\documentclass[twocolumn, amsmath, amssymb, aps]{revtex4-2}
\usepackage{graphicx}% Include figure files
\usepackage{dcolumn}% Align table columns on decimal point
\usepackage{bm}% bold math
\usepackage{booktabs}
\begin{document}

%\preprint{APS/123-QED}

\title{Anisotropic magnetic property of single crystals $R$V$_6$Sn$_6$ ($R$ = Y, Gd - Tm, Lu)}% Force line breaks with \\
%\thanks{A footnote to the article title}%

\author{Jeonghun Lee}
\author{Eundeok Mun}%
%\email{emun@sfu.ca}
\affiliation{Department of Physics, Simon Fraser University, 8888 University Dr, Burnaby, BC V5A 1S6, Canada}

%\date{\today}% It is always \today, today,
             %  but any date may be explicitly specified

\begin{abstract}
$R$V${_6}$Sn${_6}$ ($R$ =  Y, Gd - Tm, Lu) single crystals are synthesized by Sn-flux method and their physical properties are characterized by magnetization, resistivity, and specific heat measurements. Powder X-ray diffraction patterns of all samples can be well indexed with the hexagonal HfFe$_6$Ge$_6$-type structure, where rare-earth atoms form hexagonal layers and vanadium atoms form Kagome layers. At high temperatures, magnetic susceptibility measurements of moment bearing rare-earths ($R$ = Gd - Tm) follow Curie-Weiss behavior. Effective moments estimated from the polycrystalline average of magnetic susceptibility curves are consistent with the values for free $R^{3+}$ ion. Strong magnetic anisotropy due to crystalline electric field effects is observed for moment bearing rare-earths, except GdV$_6$Sn$_6$. The easy magnetization direction is determined to be $c$-axis for $R$ = Tb - Ho and $ab$-plane for $R$ = Er, and Tm. The vanadium ions in $R$V${_6}$Sn${_6}$ possess no magnetic moment. The compounds for $R$ = Y and Lu exhibit typical characteristics of paramagnetic metals. At low temperatures, the magnetic ordering is confirmed from magnetization, specific heat, and resistivity: the highest $T_{N} = 4.9$~K for GdV$_6$Sn$_6$ and the lowest $T_{N} = 2.3$~K for HoV$_6$Sn$_6$. No magnetic ordering is observed down to 1.8~K for $R$ = Er and Tm. A slight deviation of the magnetic ordering temperature from the de Gennes scaling suggests the dominant Ruderman-Kittel-Kasuya-Yosida (RKKY) exchange interaction between rare-earth moments in metallic $R$V${_6}$Sn${_6}$ compounds.
\end{abstract}

%\keywords{RV$_6$Sn$_6$\sep rare-earth magnetism\sep magnetic anisotropy\sep crystalline electric field\sep RKKY interactions}%Use showkeys class option if keyword
                              %display desired
\maketitle

\section{Introduction}
Intermetallic compounds with the chemical formula $RT_6X_6$ ($R$ = rare-earth, $T$ = V, Cr, Mn, Fe, Co,  $X$ = Ge , Sn) have shown complex magnetic and electronic properties that are emerged from the interplay between localized 4$f$ and itinerant 3$d$ electrons \cite{El1991,Venturini1991,Venturini1992,Brabers1993,Brabers1994,Cadogan2001}. The $RT_6X_6$ compounds crystallize into either the fully ordered HfFe$_6$Ge$_6$-type or other disordered YCo$_6$Ge$_6$-type or SmMn$_6$Sn$_6$-type structure, depending on the size of the constituent $R$, $T$, and $X$ atoms \cite{El1991,Venturini2006,Venturini2008,Weiland2020}. In these crystal structures the rare-earth ions form hexagonal layers and the transition metal ions form Kagome layers. When both $R$ and $T$ bear magnetic moments in their sublattices such as $R$Mn$_6$Sn$_6$, the compounds show temperature dependent magnetic structures which are governed by the relative strength of $R$-$R$, $R$-$T$, and $T$-$T$ exchange couplings \cite{Buschow1977,Venturini1992}. In $R$Mn$_6$Sn$_6$, a strong antiferromagnetic coupling between rare-earth and Mn sublattices leads to a simultaneous ordering of both sublattices \cite{Venturini1991}, where the compounds with $R$ = Gd - Ho show a collinear ferrimagnetic ordering above room temperatures and the compounds with $R$ = Er and Tm show an antiferromagnetic ordering of Mn sublattice at high temperatures followed by a ferrimagnetic ordering of $R$ sublattice at low temperatures \cite{Malaman1999}. The magnetic structure of these compounds is highly dependent on crystalline electric field (CEF) \cite{Malaman1999}. It has been observed that when the transition metals bear no magnetic moment (e.g. $T$ = Cr and Co), a very low magnetic ordering temperature arises only from the rare-earth sublattice. For example, $R$Cr$_6$Ge$_6$ compounds indicate the magnetic ordering below 15~K \cite{Mulder1993, Brabers1994} and $R$Co$_6$$X_6$ ($X$ = Ge, Sn) compounds show the magnetic ordering below 3~K \cite{Schobinger1998, Szytula2004}. In particular, GdCo$_6$Sn$_6$ is a paramagnet down to 1.8~K \cite{Szytula2004}.

In recent years, there have been considerable efforts to investigate topologically non-trivial states associated with the Kagome lattice  \cite{Ye2018, Wang2018, Yin2020}. The $R$V${_6}$Sn${_6}$ compounds have gained attention due to their topologically nontrivial band structures \cite{Ishii2013, Ghimire2020, Ma2021}, where $R$V$_6$Sn$_6$ can host 2D Kagome surface states \cite{Peng2021}. Measurements using an angle resolved photoemission spectroscopy (ARPES) on single crystals of $R$V$_6$Sn$_6$ ($R$ = Gd and Ho) and density functional theory calculations show characteristics of Dirac cone, saddle point, and flat band, which arise from the vanadium Kagome lattice \cite{Peng2021}. It is also suggested that Dirac crossing at the K-point gives rise to non-linear Hall resistivity in GdV$_6$Sn$_6$ and YV$_6$Sn$_6$ \cite{Ishikawa2021}. 

Despite growing interest in electronic properties in the $R$V${_6}$Sn${_6}$ series, their magnetic properties have not yet been studied, except for $R$ = Gd and Y. Magnetization measurements on single crystals of $R$V${_6}$Sn${_6}$ ($R$ = Gd and Y) confirm that vanadium (V) ions possess no magnetic moments and Gd ions order antiferromagnetically below $\sim$5~K \cite{Ishikawa2021, Pokharel2021}. Since the transition metal V is non-magnetic, $R$V$_6$Sn$_6$ would provide an opportunity to study magnetic properties arising solely from the rare-earth triangular lattice. It has been reported that single crystals of $R$V${_6}$Sn${_6}$ grown by the flux method adopt the HfFe$_6$Ge$_6$-type for $R$ = Gd and Y \cite{Ishikawa2021}, whereas polycrystalline samples of $R$V${_6}$Sn${_6}$ grown by the arc melt technique adopt the SmMn$_6$Sn$_6$-type for $R$ = Gd - Tm and Lu and the HfFe$_6$Ge$_6$-type for $R$ = Y \cite{Romaka2011}. In this report, we present the crystal structure and physical properties of $R$V$_6$Sn$_6$ ($R$ = Y, Gd - Tm, Lu) single crystals grown by Sn-flux.

\section{Experiments}

Single crystals of $R$V$_{6}$Sn$_{6}$ ($R$ = Y, Gd - Tm, Lu) were grown by Sn-flux \cite{Canfield1992}. The constituent elements were cut into smaller pieces and placed in an alumina crucible in the ratio $R$ : V : Sn = 1 : 6 : 62, and then sealed into an amorphous silica tube under partial Argon atmosphere. The ampoule was heated up to 1225~$^\circ$C over three hours, held there for five hours, and then slowly cooled down to 850 $^\circ$C at a rate of 1.25~$^\circ$C/hr. Single crystals were separated from the flux by centrifuging. The obtained single crystals have a plate-like morphology with a clean hexagonal faucet ($ab$-plane) being the base of the plate as shown in the inset of Fig.~\ref{Fig1} (a). All grown single crystals have similar hexagonal morphology, but compounds with heavier rare-earth elements tends to form thicker single crystals. The average thickness of the obtained single crystals varies from $\sim$0.2~mm (GdV$_{6}$Sn$_{6}$) to $\sim$1~mm (LuV$_{6}$Sn$_{6}$).

To verify the crystal structure of the title compounds and their lattice parameters, powder X-ray diffraction (XRD) patterns of crushed single crystals were collected in a Rigaku MiniFlex diffractometer at room temperature and analyzed using a {\it GSAS-II} software \cite{Toby2013}. Si standard was used to minimize the instrumental error. The dc magnetization as a function of temperature from 1.8 to 300 K and magnetic fields up to 70 kOe was measured in a Quantum Design (QD) Magnetic Property Measurement System (MPMS). Four-probe ac resistivity measurements were performed with $I\parallel ab$ from 300~K down to 1.8~K in a QD Physical Property Measurement System (PPMS). Specific heat was measured by the relaxation technique down to 1.8~K in a QD PPMS.

\section{Results and discussion}

\subsection{crystal structure}

\begin{figure}
\centering
\includegraphics[width=1\linewidth]{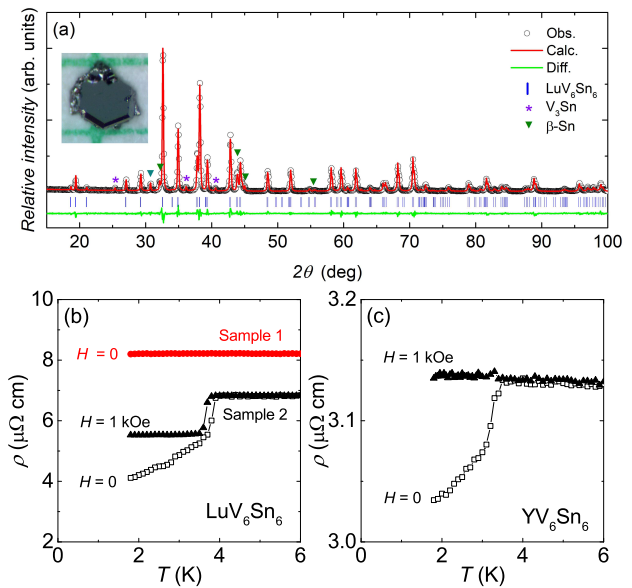}
\caption{(a) Powder XRD patterns of LuV$_6$Sn$_6$. Red line is the calculated XRD patterns using HfFe$_6$Ge$_6$-type structure. Green line represents the difference between the data and calculated patterns. Impurity phases of V$_3$Sn and $\beta$-Sn are marked by purple asterisk and green triangle symbols, respectively. Inset shows a photograph of LuV$_6$Sn$_6$ on mm grid. (b) Electrical resistivity of two LuV$_6$Sn$_6$ samples at $H$ = 0 and 1~kOe. (c) Electrical resistivity of YV$_6$Sn$_6$ at $H$ = 0 and 1~kOe.}
\label{Fig1}
\end{figure}

The obtained and calculated powder XRD patterns for LuV$_6$Sn$_6$, as a representative example in the series, are shown in Fig.~\ref{Fig1}~(a). The observed peak positions are well indexed by the fully ordered HfFe$_6$Ge$_6$-type structure ($P6/mmm$, no. 191). In earlier studies the two structural prototypes for $R$V$_6$Sn$_6$ compounds are reported: single crystalline samples (grown by Sn-flux) for $R$ = Gd and Y \cite{Ishikawa2021} crystallize into the fully occupied HfFe$_6$Ge$_6$-type; polycrystalline samples (synthesized by arc melt) crystallize into the HfFe$_6$Ge$_6$-type for $R$ = Y and the SmMn$_6$Sn$_6$-type for $R$ = Gd - Tm \cite{Romaka2011}. The SmMn$_6$Sn$_6$-type is a disordered variant of the HfFe$_6$Ge$_6$-type. In the SmMn$_6$Sn$_6$-type, the fractional occupancy of additional $R$ and Sn sites are reported to vary from 3 to 10~\%, depending on $R$ \cite{Romaka2011}. In this study, the fully ordered HfFe$_6$Ge$_6$-type structure is used to analyze the powder XRD patterns as it does not show any appreciable difference between the observed and calculated intensities, as shown in Fig.~\ref{Fig1}~(a). The detection of subtle difference between the two structure types requires a much higher resolution diffractometer with rigorous Rietveld refinement.

\begin{figure}
\centering
\includegraphics[width=1\linewidth]{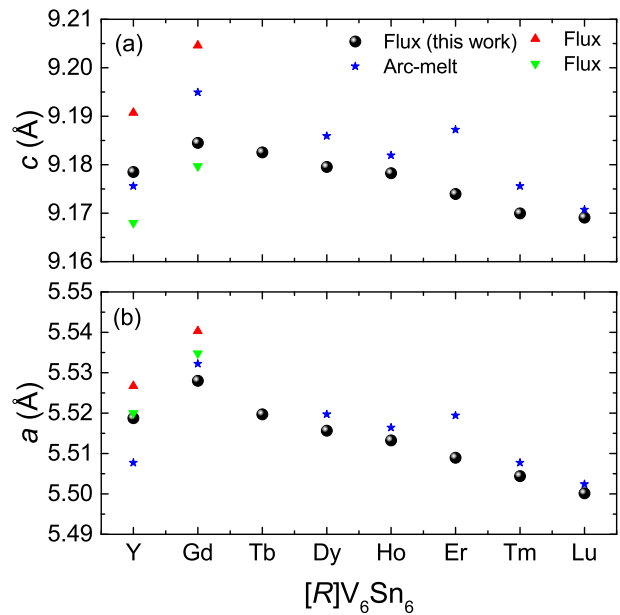}
\caption{(a) Lattice parameter $c$ and (b) lattice parameter $a$ as a function of rare-earth elements [$R$] (black circles). Blue stars \cite{Romaka2011}, red up-triangles \cite{Ishikawa2021}, and green down-triangles \cite{Pokharel2021} are taken from other reports.}
\label{Fig2}
\end{figure}

The obtained lattice parameters $a$ and $c$ are shown in Fig.~\ref{Fig2} (a) and (b), respectively. Both $a$ and $c$ decrease with increasing $R$ atomic number, following lanthanide contraction, implying that the valence state of the heavy rare-earth ions in $R$V$_6$Sn$_6$ compounds are trivalent. The lattice parameters obtained from earlier studies \cite{Ishikawa2021, Pokharel2021,Romaka2011} are included in Fig.~\ref{Fig2}. The variations in lattice parameters are probably due to the different sample quality, refinement with different structure-type, and instrumental error.

\subsection{impurity phase}

The obtained powder XRD patterns and physical property measurements performed on this series show two minor impurity phases. As shown in Fig.~\ref{Fig1}~(a), a traceable amount of V$_3$Sn and $\beta$-Sn phases are detected in LuV$_6$Sn$_6$. These two are the only impurity phases detected in the $R$V$_6$Sn$_6$ series and also consistent with impurities found in the previous report \cite{Romaka2011}. V$_3$Sn and $\beta$-Sn phases are known to have superconducting transitions at $T_c$ = 3.8~K \cite{Hatt1973} and 3.7~K \cite{Tian2003}, respectively. The presence of these superconducting phases are clearly seen in electrical resistivity, $\rho(T)$, measurements. $\rho(T)$ curves on two different LuV$_6$Sn$_6$ samples are shown in Fig.~\ref{Fig1}~(b). Sample 1 and 2 are two different pieces of single crystals grown from the same batch. $\rho(T)$ of the sample 1 indicates no resistivity drop below 4~K. $\rho(T)$ of the sample 2 shows the superconducting transition below 4~K at $H$ = 0, which cannot be fully suppressed by applying magnetic field of 1~kOe. This is probably due to the superconducting V$_3$Sn phase. For the YV$_6$Sn$_6$ sample, its partial superconducting transition can be fully suppressed by a magnetic field of 1~kOe, as shown in Fig.~\ref{Fig1}~(c). The superconducting transition in YV$_6$Sn$_6$ probably originates from $\beta$-Sn. It should be emphasized that superconducting transitions observed in resistivity measurements are not intrinsic to the ternary compound, as its resistivity does not go to zero below $T_c$. Thus, the superconducting transition below 4~K originates from impurity phases either on surfaces or embedded within the samples.

\subsection{LuV$_{6}$Sn$_{6}$ and YV$_{6}$Sn$_{6}$}

\begin{figure}
\centering
\includegraphics[width=1\linewidth]{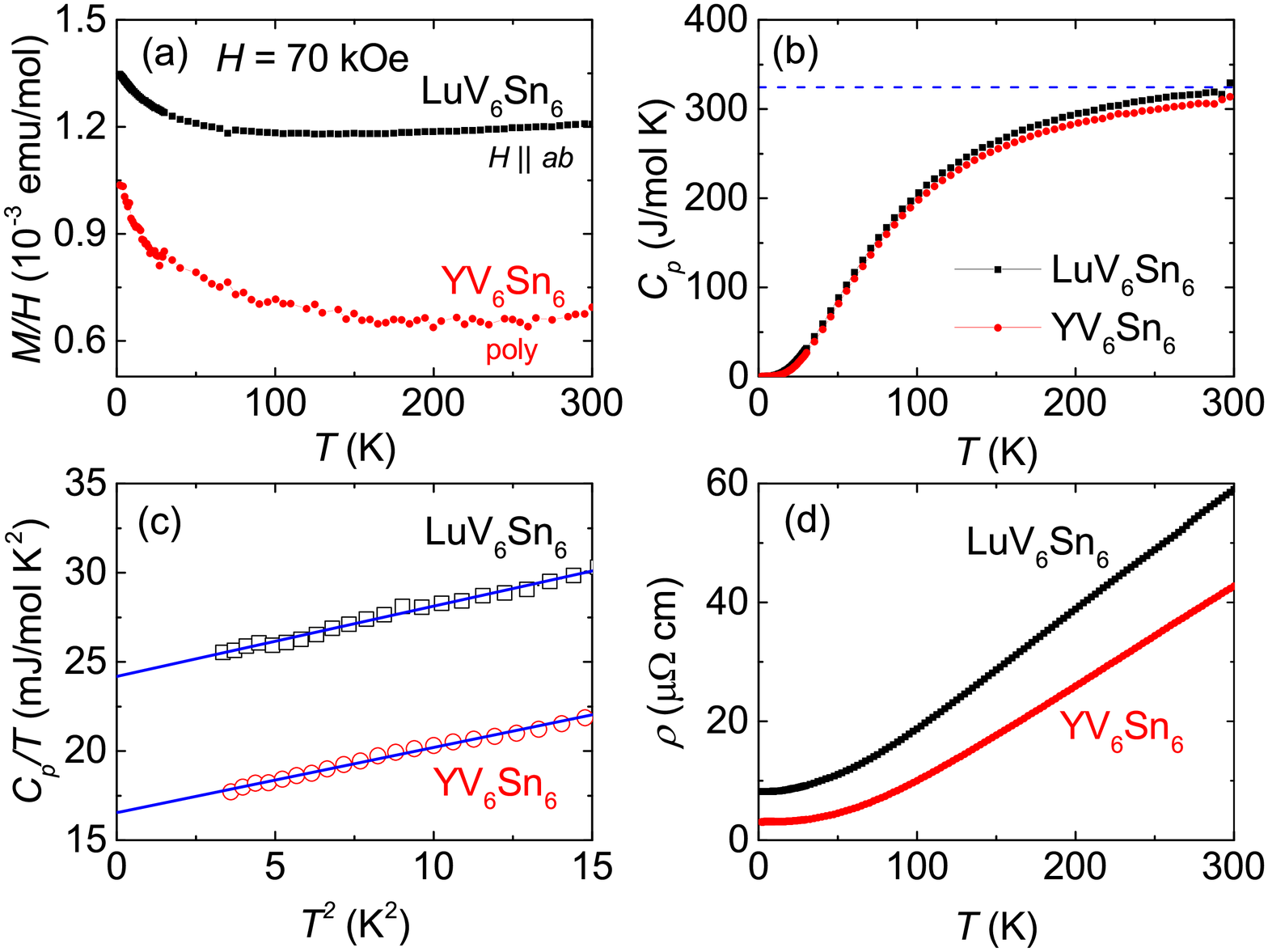}
\caption{Physical property measurements of LuV$_6$Sn$_6$ and YV$_6$Sn$_6$. (a) $M/H$ at $H$ = 70~kOe. (b) Specific heat at $H$~=~0. Horizontal line represents the Dulong-Petit limit. (c) $C_{p}/T$ vs. $T^2$ plot for LuV$_6$Sn$_6$ at $H$ = 20~kOe and YV$_6$Sn$_6$ at $H$ = 30~kOe. Solid lines represent fits to $C_{p}(T) = \gamma T + \beta T^{3}$. (d) Electrical resistivity at $H$ = 0.}
\label{Fig3}
\end{figure}

The temperature-dependent magnetic susceptibility, $\chi = M/H$, curves for $R$ = Lu and Y at $H$~=~70~kOe are shown in Fig.~\ref{Fig3} (a). Due to the small size of single crystals, multiple pieces of YV$_6$Sn$_6$ were loaded into a gel capsule to measure magnetic susceptibility, which can be considered as a polycrystalline average. At high temperatures $M/H$ curves of both samples are weakly dependent on temperature, while $M/H$ curves at low temperatures exhibit an upturn probably due to paramagnetic impurities. The absolute value of magnetic susceptibility of both compounds is a bit larger than that of typical Y- and Lu-based intermetallic compounds \cite{Sefat2008, Myers1999, Bud1999}. For YV$_{6}$Sn$_{6}$ the absolute value of $M/H$ at 300~K has the similar order of magnitude as previously reported values ($\sim$10$^{-3}$ emu/mol) \cite{Ishikawa2021,Pokharel2021}. Thus, the enhanced magnetic susceptibility seems to be the generic feature in non-magnetic $R$V$_6$Sn$_6$ compounds. From the magnetization measurements we infer that the V ions in these compounds do not carry magnetic moment. Typically, a magnetic moment on V is observed in vanadium complexes such as insulating spinel ZnV$_2$O$_4$ (V$^{3+}$, $S$ = 1) \cite{Reehuis2003} and triangular lattice antiferromagnet VCl$_2$ (V$^{2+}$, $S$ = 3/2) \cite{Kadowaki1987}. The absence of the magnetic moment on V ions in $R$V$_6$Sn$_6$ implies quenching of the spin magnetic moment \cite{Blades2017}, where Sn atoms may have contributed to quenching of the spin moment on vanadium by forming metallic or covalent bonds, as evidenced by short V-Sn distances \cite{Romaka2011}. 

The temperature-dependent specific heat, $C_{p}(T)$, curves of LuV$_{6}$Sn$_{6}$ and YV$_{6}$Sn$_{6}$ are shown Fig.~\ref{Fig3} (b). At high temperatures the $C_{p}(T)$ curves of both compounds are almost identical and approach the Dulong-Petit limit at 300~K. At low temperatures, $C_{p}(T)$ can be described by considering electronic and phonon contributions: $C_{p}(T) = \gamma T + \beta T^3$. Figure~\ref{Fig3} (c) shows plots of $C/T$ vs. $T^2$ for $R$ = Lu at $H$ = 20~kOe and $R$ = Y at 30~kOe. Due to the superconducting transition below 4~K the specific heat data under magnetic field are used to estimate the electronic specific heat coefficient ($\gamma$) and Debye temperature ($\theta_D$). The obtained $\gamma$ values are $\sim$24~mJ/mole~K$^{2}$ for $R$ = Lu and $\sim$17~mJ/mole K$^{2}$ for $R$ = Y. The estimated $\theta_D$ from $\beta$ is $\sim$400~K for both compounds. Note that a rather high $\gamma$ value ($\sim$67~mJ/mole K$^{2}$), obtained by fitting $C/T$ vs. $T^2$ curve from 200 to 600~K$^2$, has been reported for YV$_{6}$Sn$_{6}$ sample  \cite{Pokharel2021}. The $\rho(T)$ curves for both $R$ = Lu and Y are shown in Fig.~\ref{Fig3} (d). The resistivity curves decrease monotonically as temperature is lowered down to 1.8 K, following a typical metallic behavior.

\subsection{Physical properties of $R$V$_6$Sn$_6$ ($R$ = Gd - Tm)}

\begin{table*}%[width=0.85\textwidth,cols=10,pos=h]
\caption{\label{tab:table1} A summary of magnetic properties of $R$V$_6$Sn$_6$ ($R$ = Gd - Tm): magnetic ordering temperatures $T_{N}^{\chi}$ determined from $d\chi T/dT$ at $H$ = 1~kOe, $T_{N}^{\rho}$ determined from $d\rho/dT$ at $H$ = 0, $T_{N}^{C_m}$ determined from $C_m$ at $H$ = 0; easy magnetization direction direction; Weiss temperatures $\theta_{p}^{ab}$, $\theta_{p}^{c}$, and $\theta_{p}^{poly}$; effective moment $\mu_{eff}$ (theoretical value for free $R^{3+}$ ion); and CEF parameter $B^{0}_{2}$.}
\begin{tabular}{@{}cccccrrrcc@{}}%{ccccccccc}
\toprule
$R$&$T_{N}^{\chi}$(K)&$T_{N}^{\rho}$(K)&$T_{N}^{C_m}$(K)&easy-direction&$\theta_{p}^{ab}$ (K)&$\theta_{p}^{c}$ (K)&$\theta_{p}^{poly}$ (K)&$\mu_{eff}$ ($\mu_{B}/R^{3+}$)&$B^{0}_{2}$(K)\\
\midrule
Gd & 4.8 &5.0&4.9& & 0.9 & -2.5 & -0.2 & 8.1 (7.9)& 0.2 \\
Tb & 4.3 &4.2&4.2& $c$-axis & -38.3 & 32.3 & -14.3 & 10.0 (9.7)& -1.4\\
Dy & 2.9 &2.9&3.0& $c$-axis & -13.4 & 17.0 & -3.5 & 10.5 (10.7)& -0.4\\
Ho & 2.3 &2.3&2.4& $c$-axis & -4.5 & 4.5 & -1.5 & 10.9 (10.6)& -0.1\\
Er & &&& $ab$-plane & 3.3 & -29.9 & -7.8 & 9.9 (9.6)& 0.4\\
Tm & &&& $ab$-plane & 10.3 & -71.2 & -17.2 & 8.1 (7.6)& 1.7\\
\bottomrule
\end{tabular}
\end{table*}

\begin{figure}
\centering
\includegraphics[width=1\linewidth]{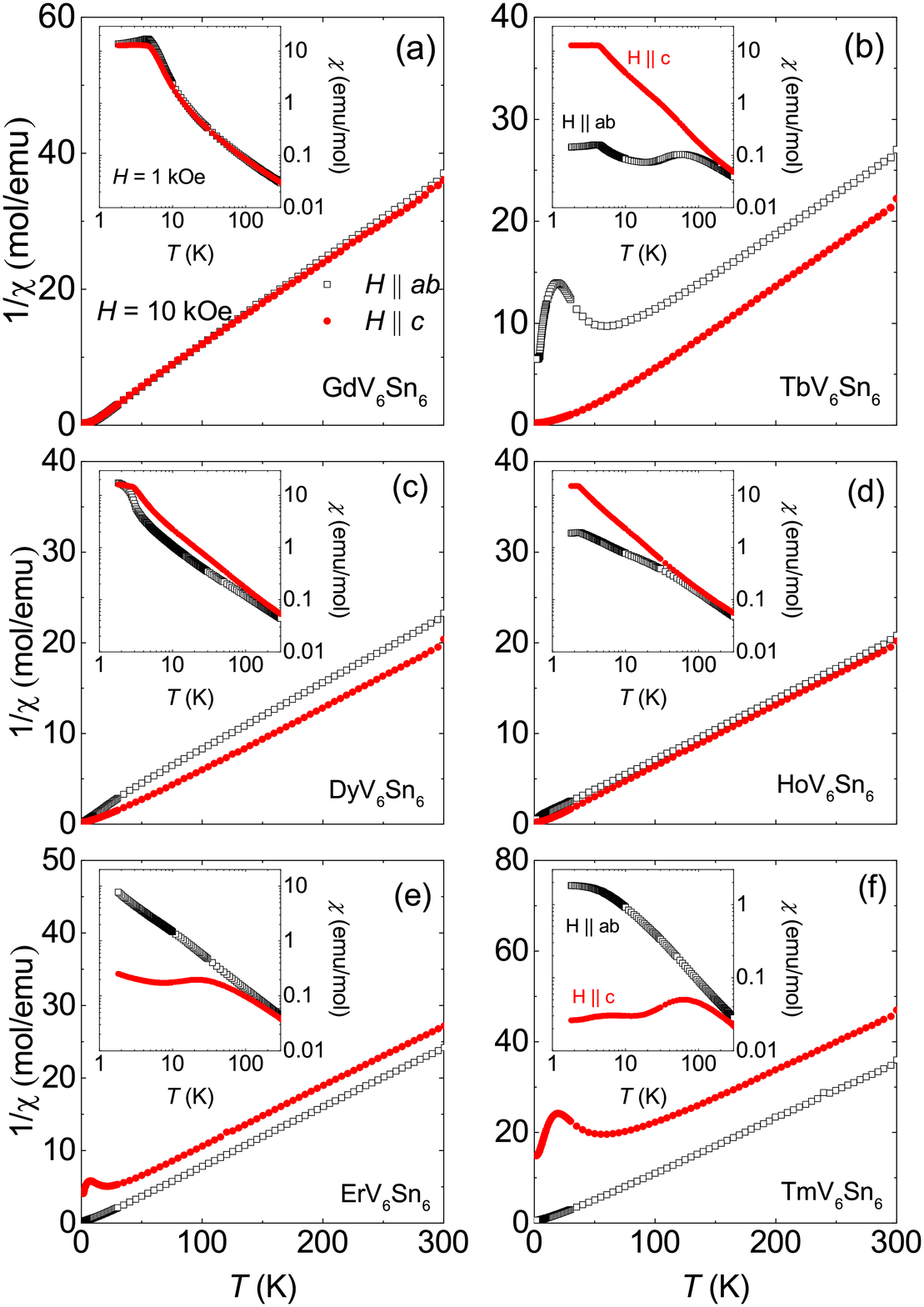}
\caption{Inverse magnetic susceptibility curves of $R$V$_6$Sn$_6$ ($R$ = Gd - Tm) at $H$ = 10~kOe for $H \parallel ab$ (open squares) and $H \parallel c$ (closed circles). Insets show $M/H$ curves at $H$~=~1~kOe.}
\label{Fig4}
\end{figure}

The magnetic susceptibility of all moment bearing members in $R$V$_6$Sn$_6$ follows the Curie-Weiss, $\chi(T) = C/(T - \theta_{p})$, behavior at high temperatures. The inverse magnetic susceptibility, $1/\chi = H/M$, curves of $R$V$_6$Sn$_6$ are plotted in Fig.~\ref{Fig4} for both $H \parallel ab$ and $H \parallel c$ at $H$ = 10~kOe. A polycrystalline average is estimated by $\chi_{poly} = 2/3 \chi_{ab} + 1/3 \chi_{c}$. Effective moments, $\mu_{eff}$, and Weiss temperatures, $\theta_{p}$, are obtained by fitting $1/\chi$ curves above 150~K to the Curie-Weiss law. The effective moments of $R$V$_6$Sn$_6$ estimated from polycrystalline average are close to that of free $R^{3+}$-ion values, as summarized in Table~\ref{tab:table1}, implying a 3+ valence state of the rare-earth ions ($R$ = Gd - Tm) and no magnetic moment on V ions. Thus, magnetic properties of $R$V$_6$Sn$_6$ can be explained by considering $4f$ moments of rare-earth ions. The $\theta_{p}^{poly}$ values obtained from polycrystalline average are negative for all rare-earths, indicating antiferromagnetic exchange interactions between rare-earth moments. For GdV$_6$Sn$_6$, $\mu_{eff}$ = 8.1~$\mu_{B}$  and $\theta_{p}$ = -0.2~K, which is consistent with that of previous polycrystalline study \cite{Ishikawa2021}. However, $\theta_{p}$ of earlier single crystal study is positive ($\sim$7.6~K) \cite{Pokharel2021}, which is somewhat larger than our result and other report \cite{Ishikawa2021}. This discrepancy may be related to different growth conditions.

%Magnetic ordering
The $M/H$ curves of $R$V$_{6}$Sn$_{6}$, measured at $H$ = 1~kOe for both $H\parallel ab$ and $H\parallel c$, are shown in the insets of Fig.~\ref{Fig4}. The $M/H$ curves for $R$ = Gd - Ho show slope changes below 5~K as a signature of the antiferromagnetic ordering. The magnetic ordering temperature determined from $d\chi T/dT$ is $T_N$ = 4.8, 4.3, 2.9, and 2.3~K for $R$ = Gd, Tb, Dy, and Ho, respectively. The magnetic ordering is not detected down to 1.8~K for $R$ = Er and Tm.

%Magnetization isotherms

\begin{figure}
\centering
\includegraphics[width=0.5\linewidth]{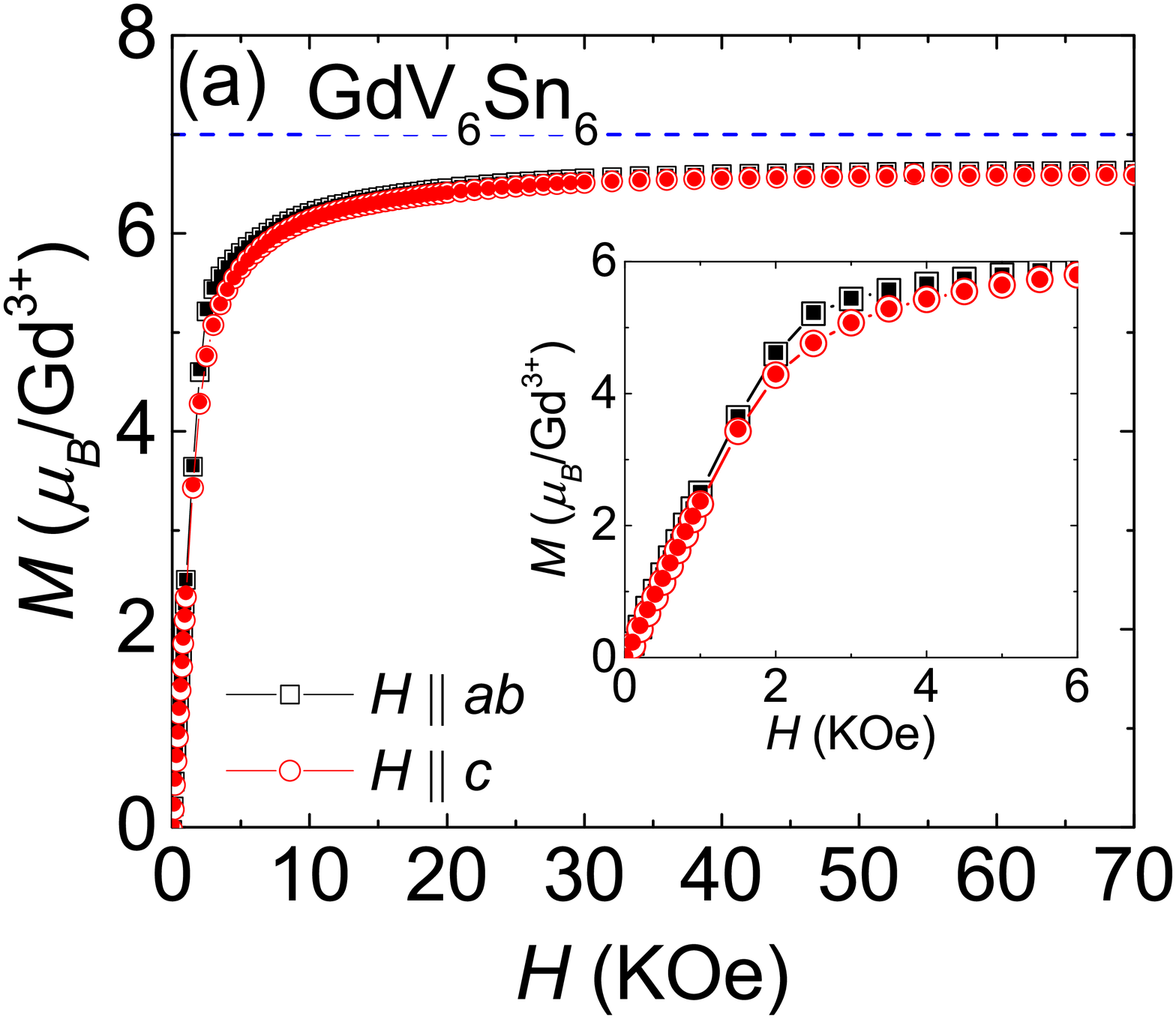}\includegraphics[width=0.5\linewidth]{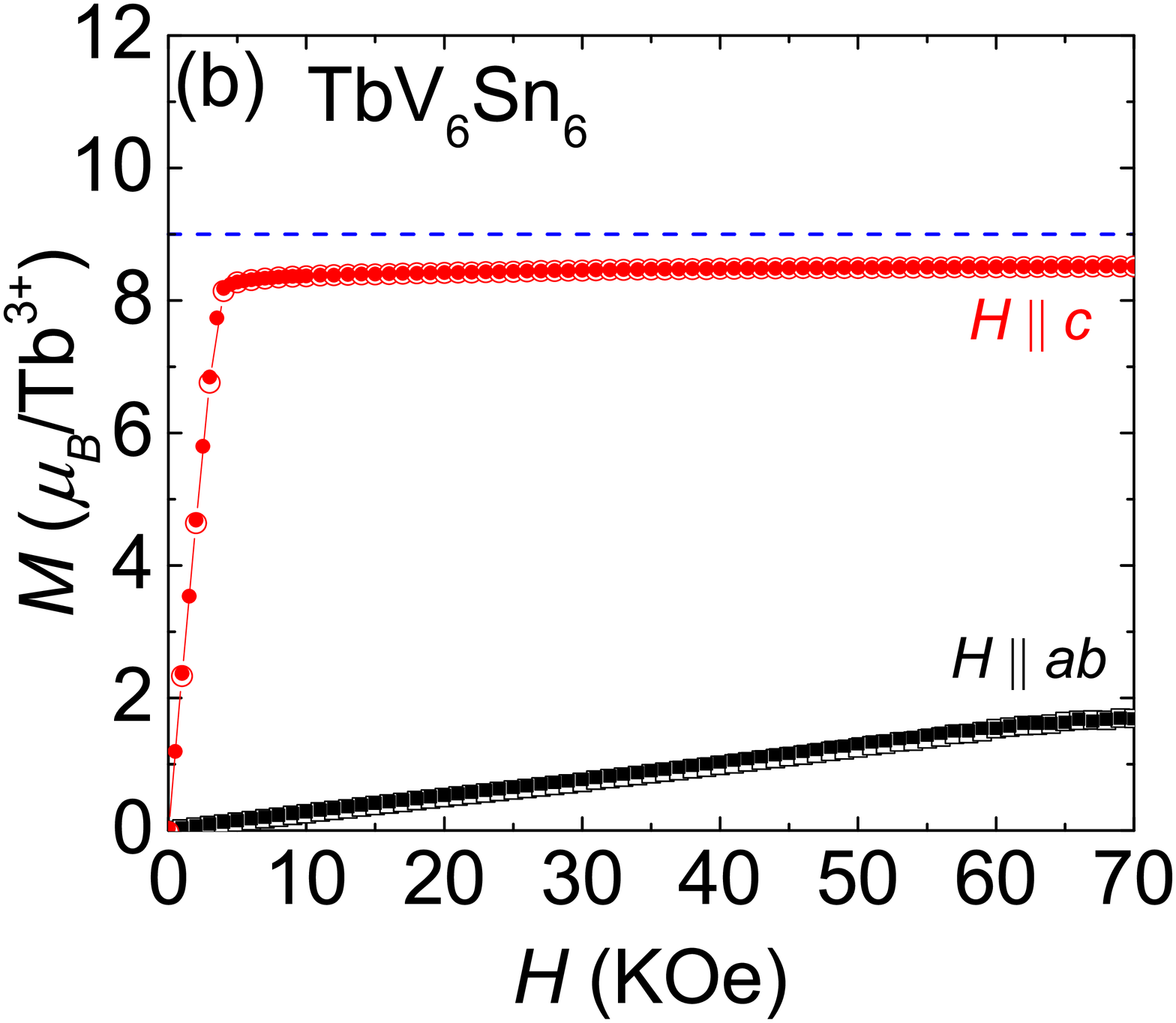}
\includegraphics[width=0.5\linewidth]{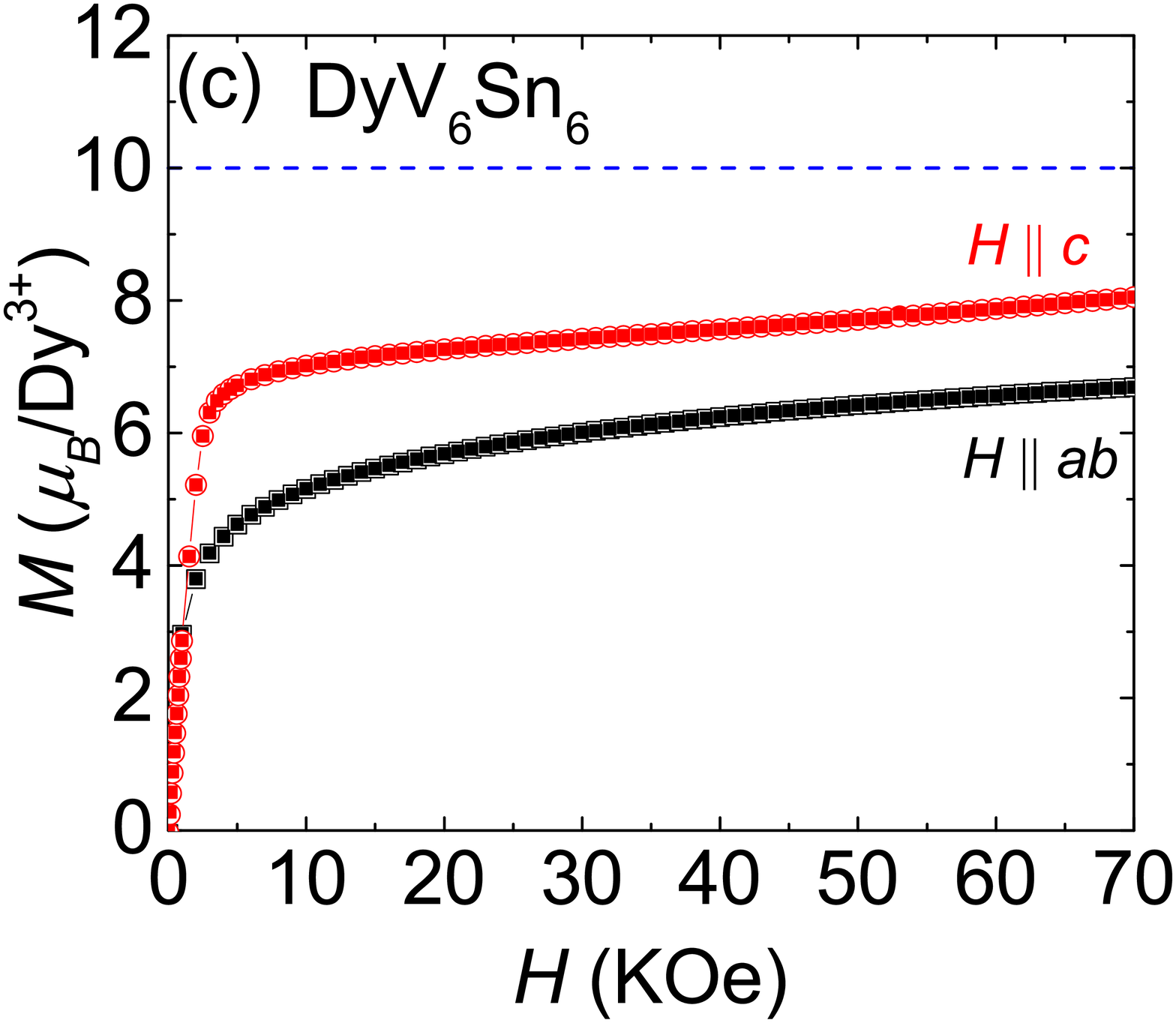}\includegraphics[width=0.5\linewidth]{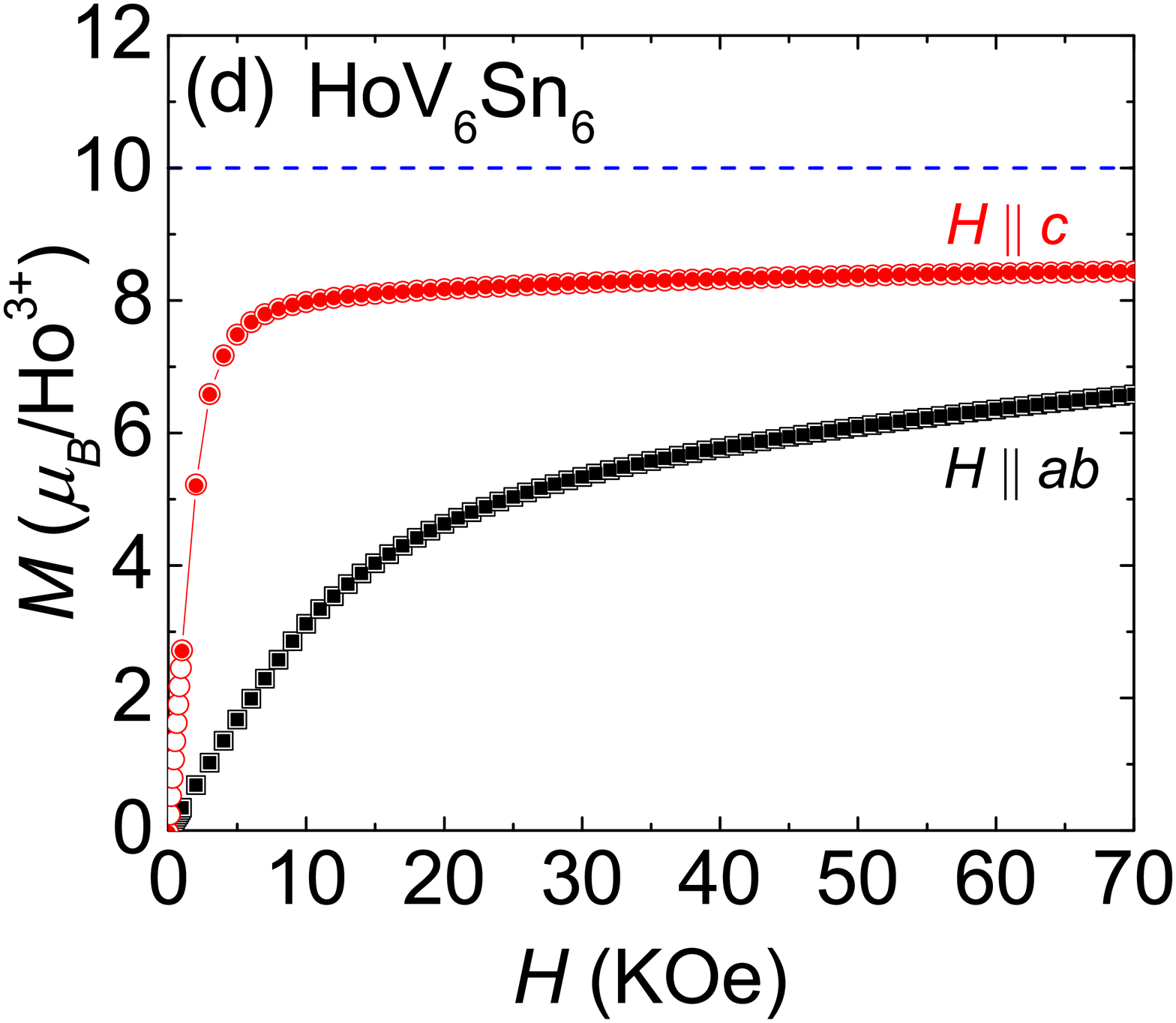}
\includegraphics[width=0.5\linewidth]{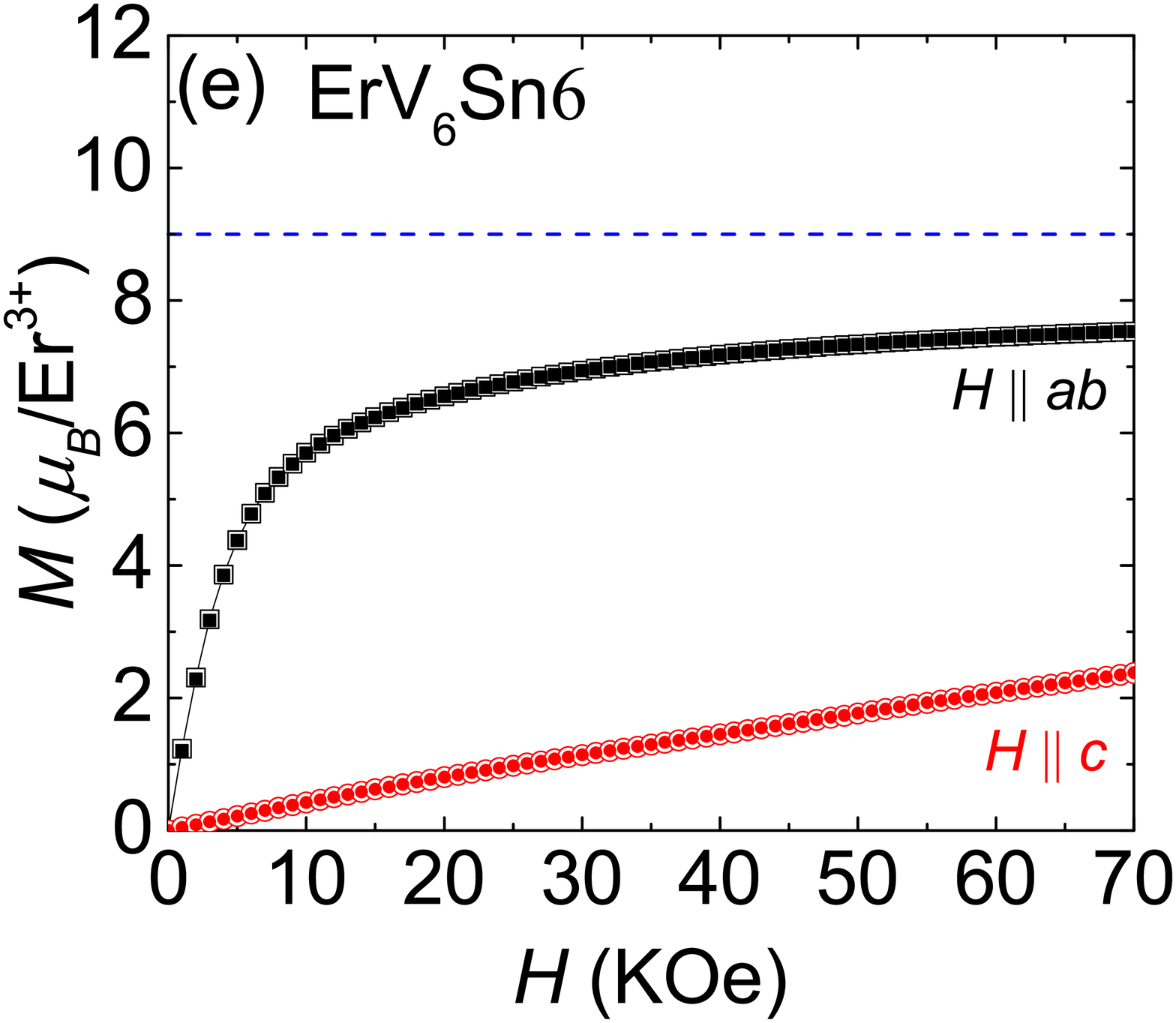}\includegraphics[width=0.5\linewidth]{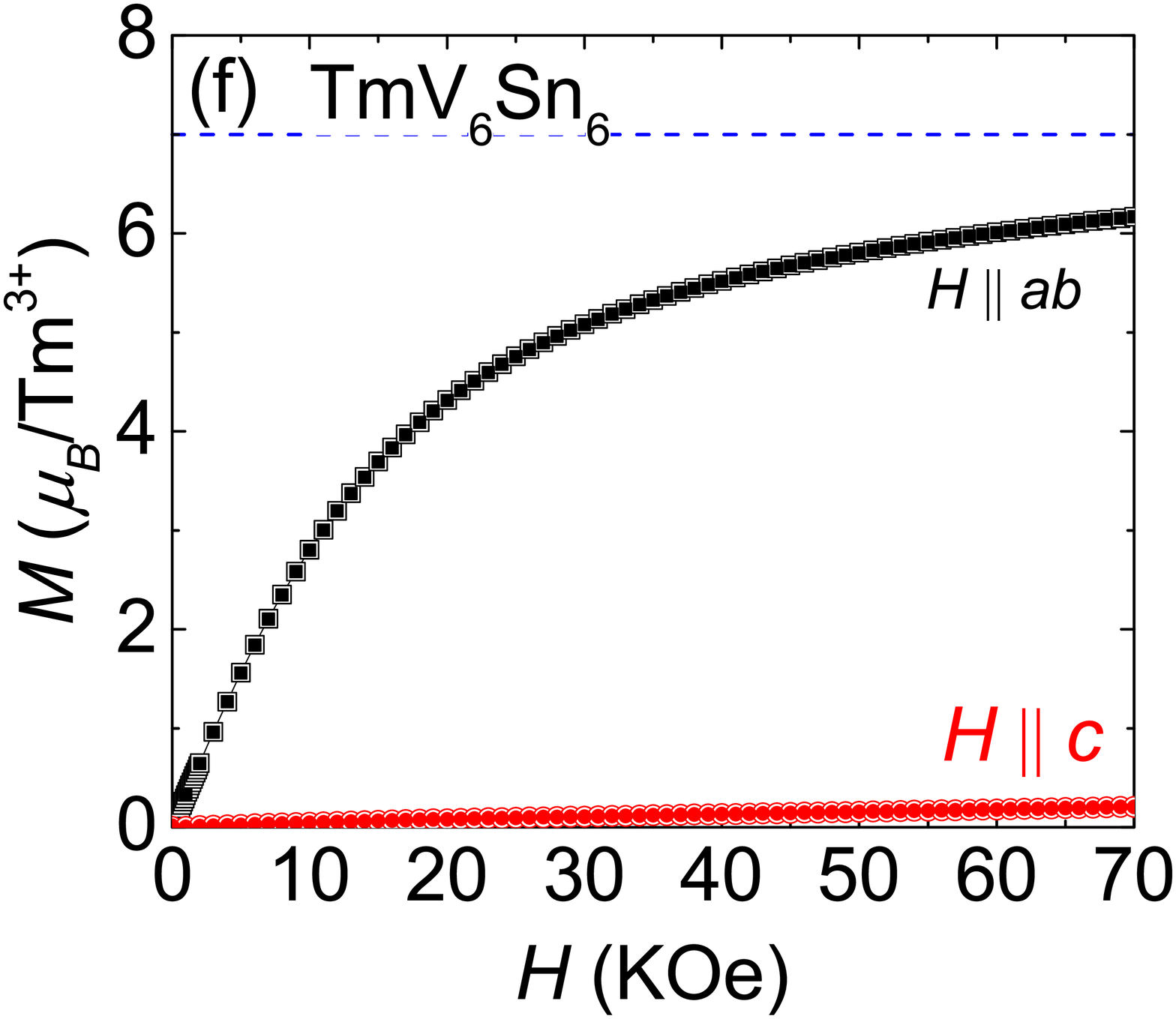}
\caption{Magnetization isotherms of $R$V$_6$Sn$_6$ ($R$ = Gd - Tm) at $T$ = 2~K for $H \parallel ab$ (squares) and $H \parallel c$ (circles). Open and closed symbols are data taken while increasing and decreasing magnetic fields, respectively. Horizontal dotted lines indicate the saturated magnetization values ($gJ$) for free $R^{3+}$ ions. Inset in (a) shows an expanded plot between 0 and 6 kOe.}
\label{Fig5}
\end{figure}

The magnetization isotherm, $M(H)$, at $T$ = 2~K in this series clearly shows a magnetic anisotropy between $H \parallel ab$ and $H \parallel c$, as shown in Fig.~\ref{Fig5}. Magnetization measurements indicate no detectable hysteresis for all $R$V$_6$Sn$_6$. The easy magnetization direction is along the $c$-axis for $R$ = Tb, Dy, and Ho and $ab$-plane for $R$ = Er and Tm. As expected, $M(H)$ of GdV$_6$Sn$_6$ indicates no anisotropy at high magnetic fields, but $M(H)$ for $H\parallel c$ is slightly smaller than that for $H\parallel ab$, as displayed in the inset of Fig.~\ref{Fig5} (a). The saturated magnetization values of GdV$_6$Sn$_6$ and TbV$_6$Sn$_6$ at 70~kOe are close to the theoretical values of free Gd (7~$\mu_{B}$/Gd) and Tb (9~$\mu_{B}$/Tb) ion. $M(H)$ of GdV$_6$Sn$_6$ is similar to ones observed in previous reports \cite{Ishikawa2021, Pokharel2021}. The saturated magnetization values for $R$ = Dy - Tm at 70~kOe are somewhat smaller than their theoretical $gJ$ values. The observed large magnetic anisotropy implies the presence of the strong CEF acting on 4$f$ moments.

%electrical resistivity

\begin{figure}
\centering
\includegraphics[width=0.5\linewidth]{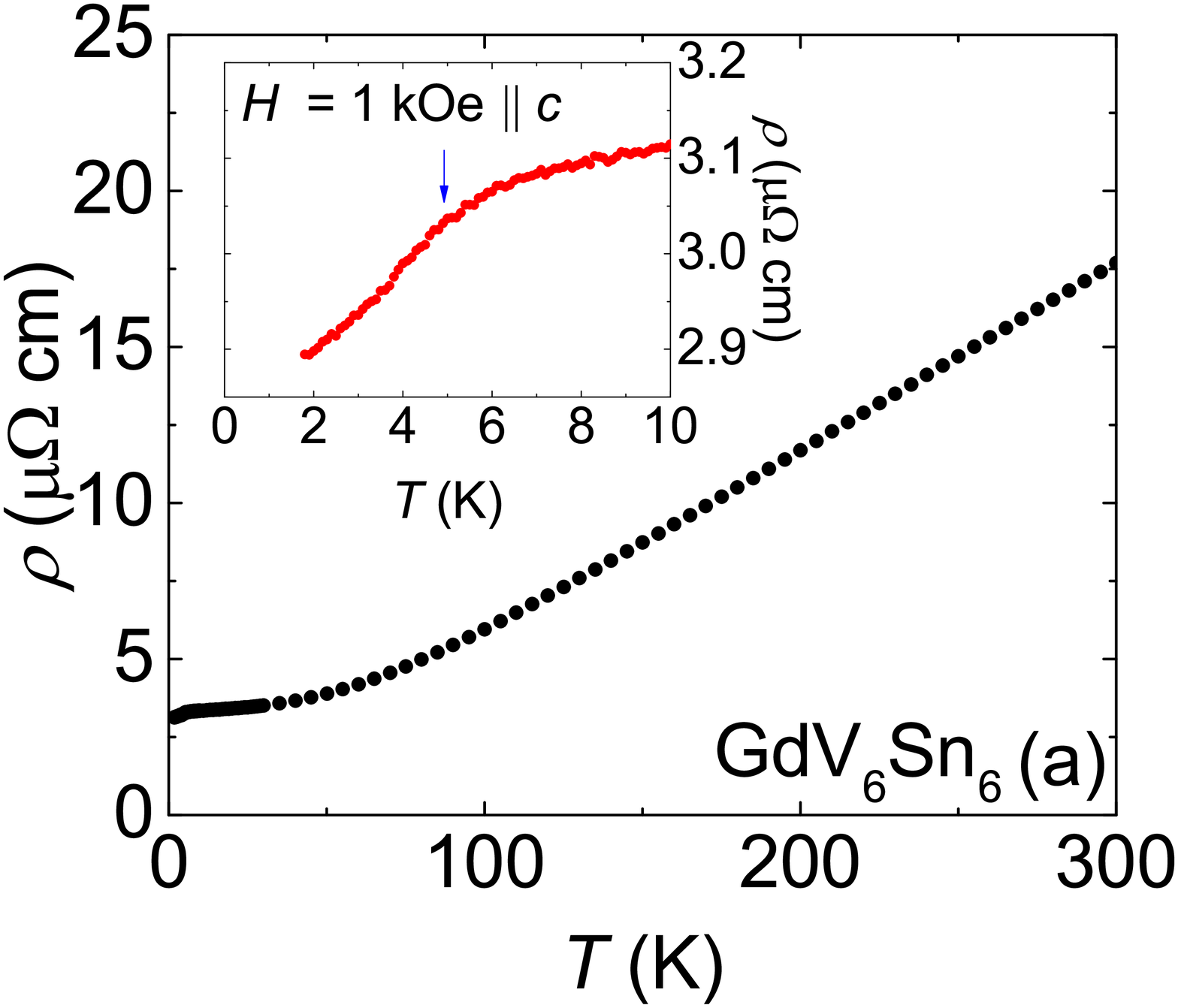}\includegraphics[width=0.5\linewidth]{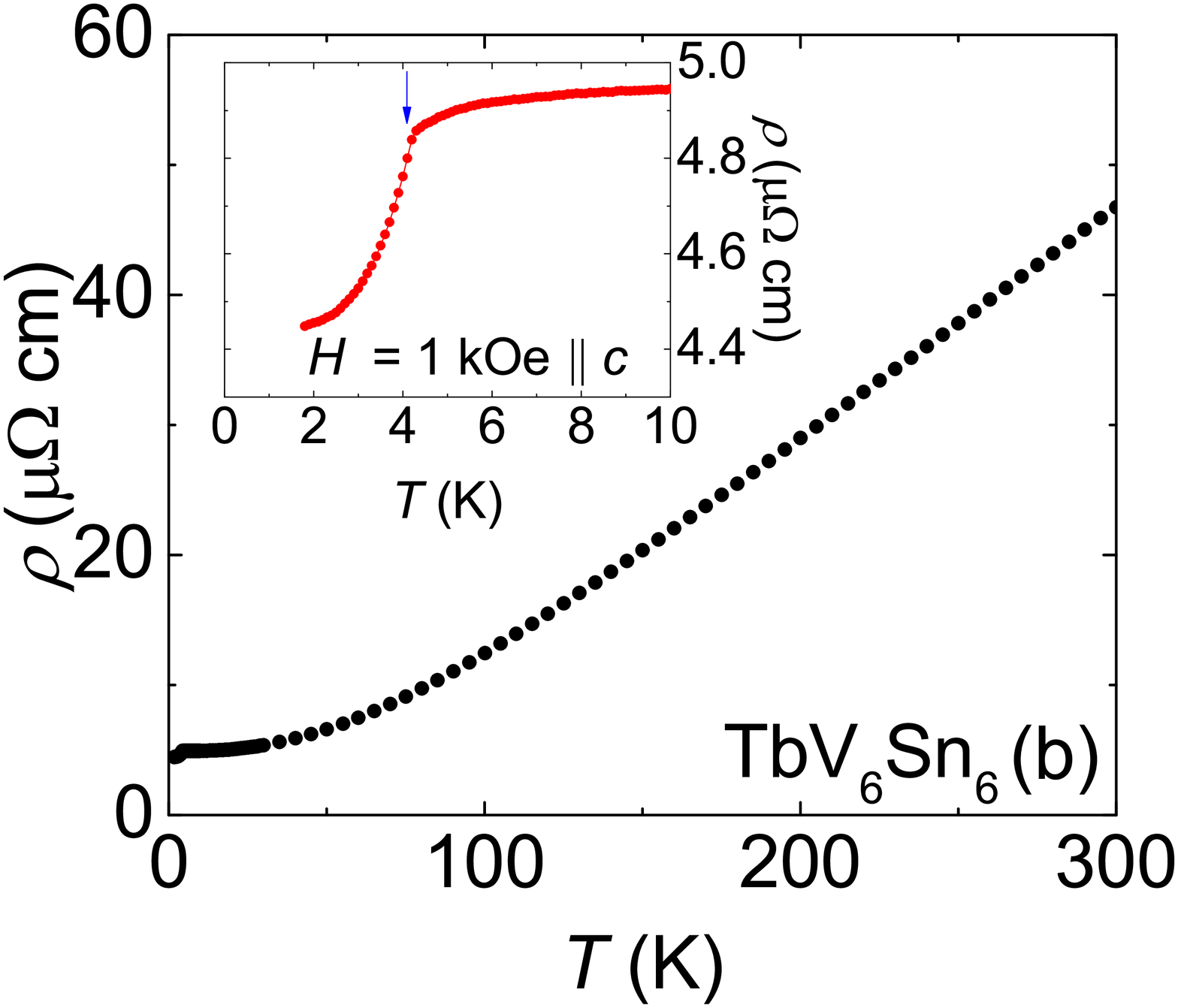}
\includegraphics[width=0.5\linewidth]{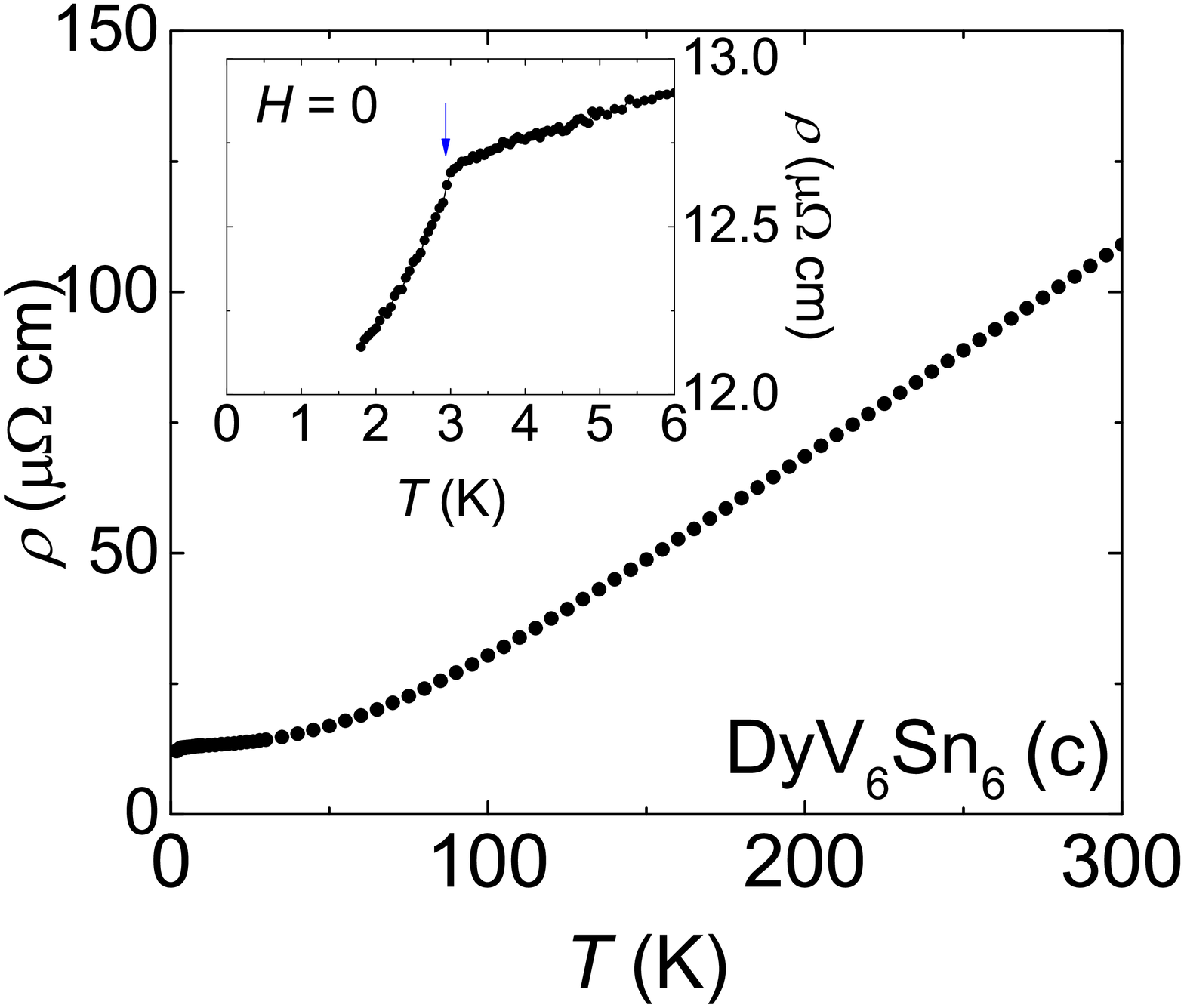}\includegraphics[width=0.5\linewidth]{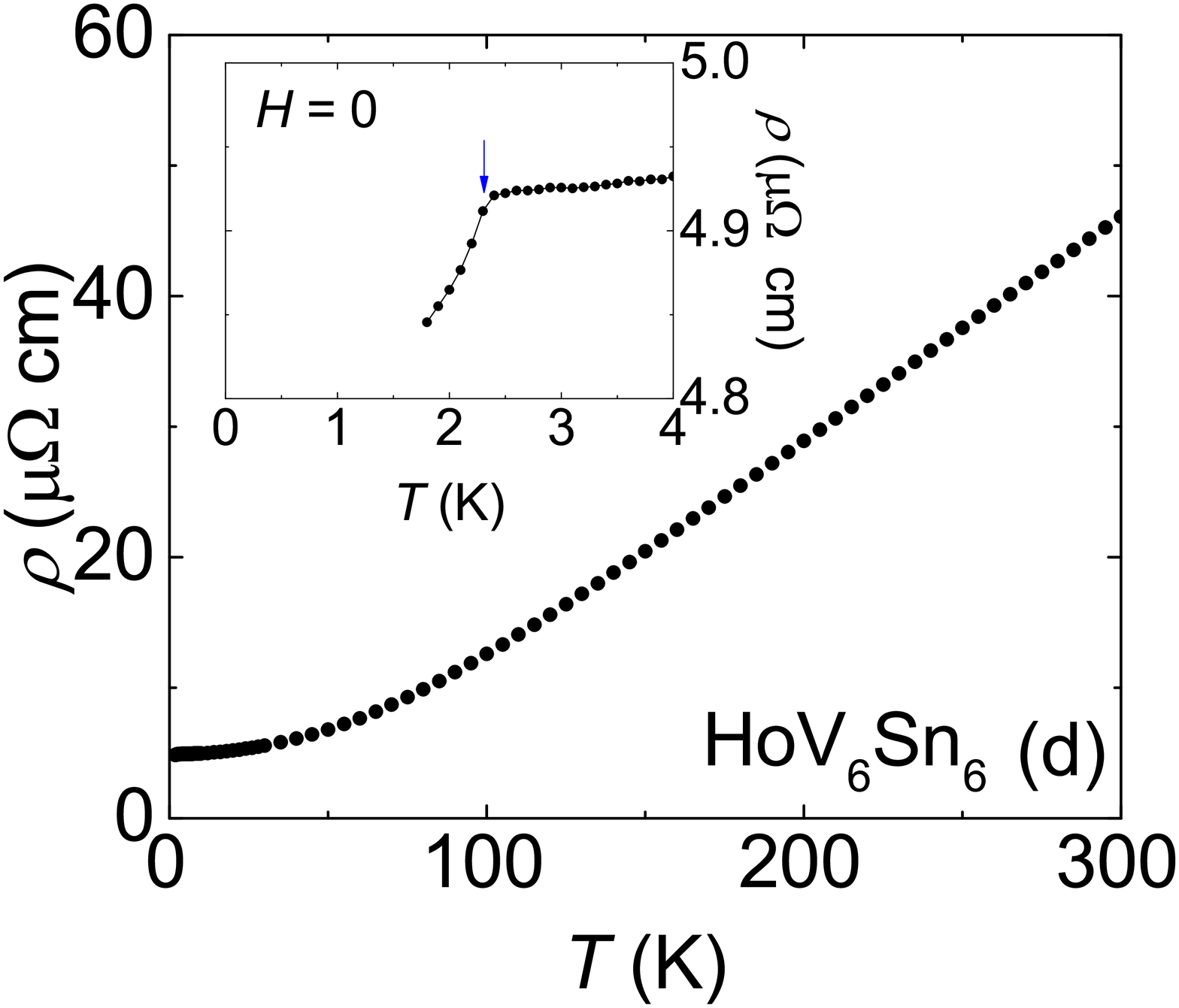}
\includegraphics[width=0.5\linewidth]{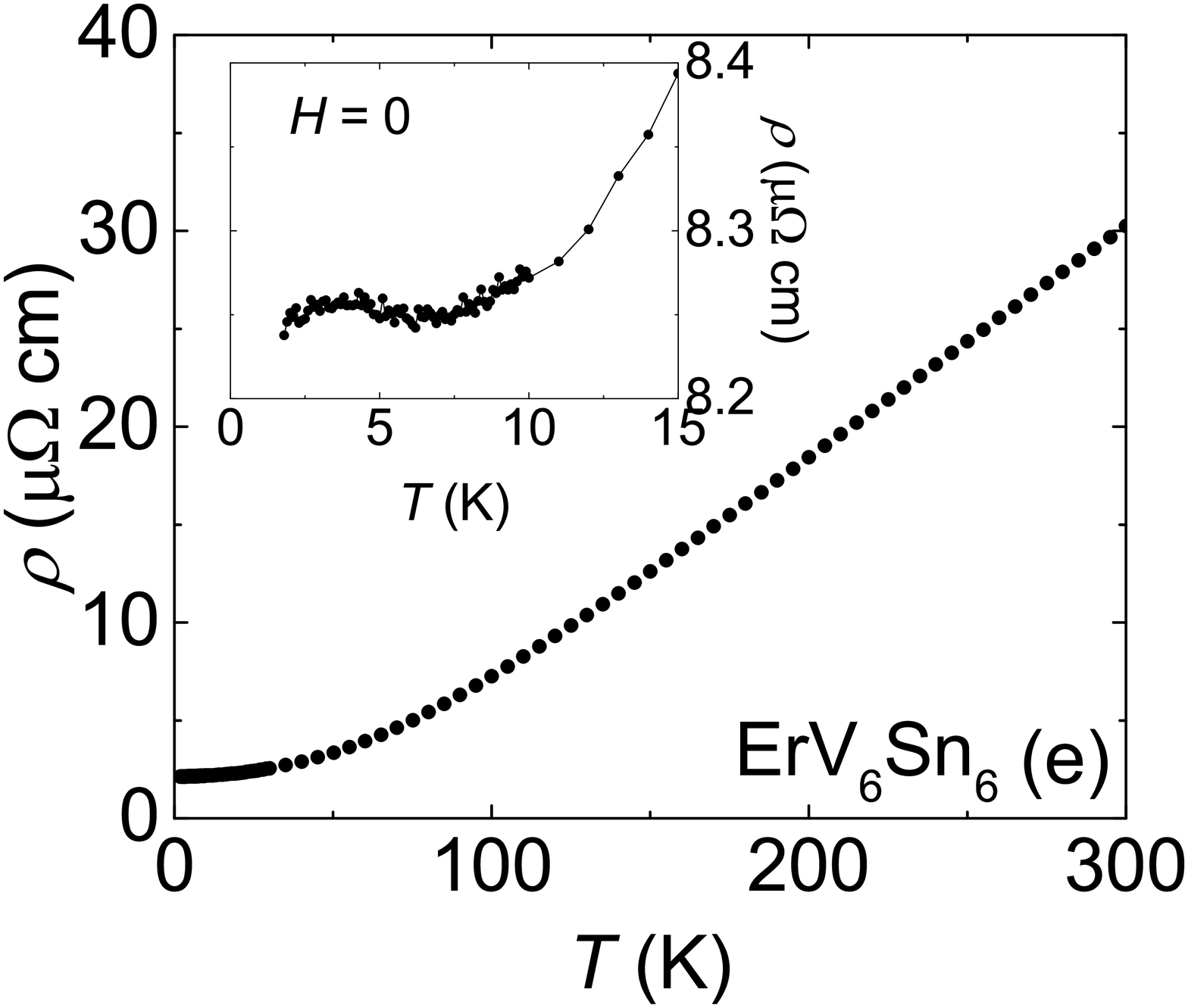}\includegraphics[width=0.5\linewidth]{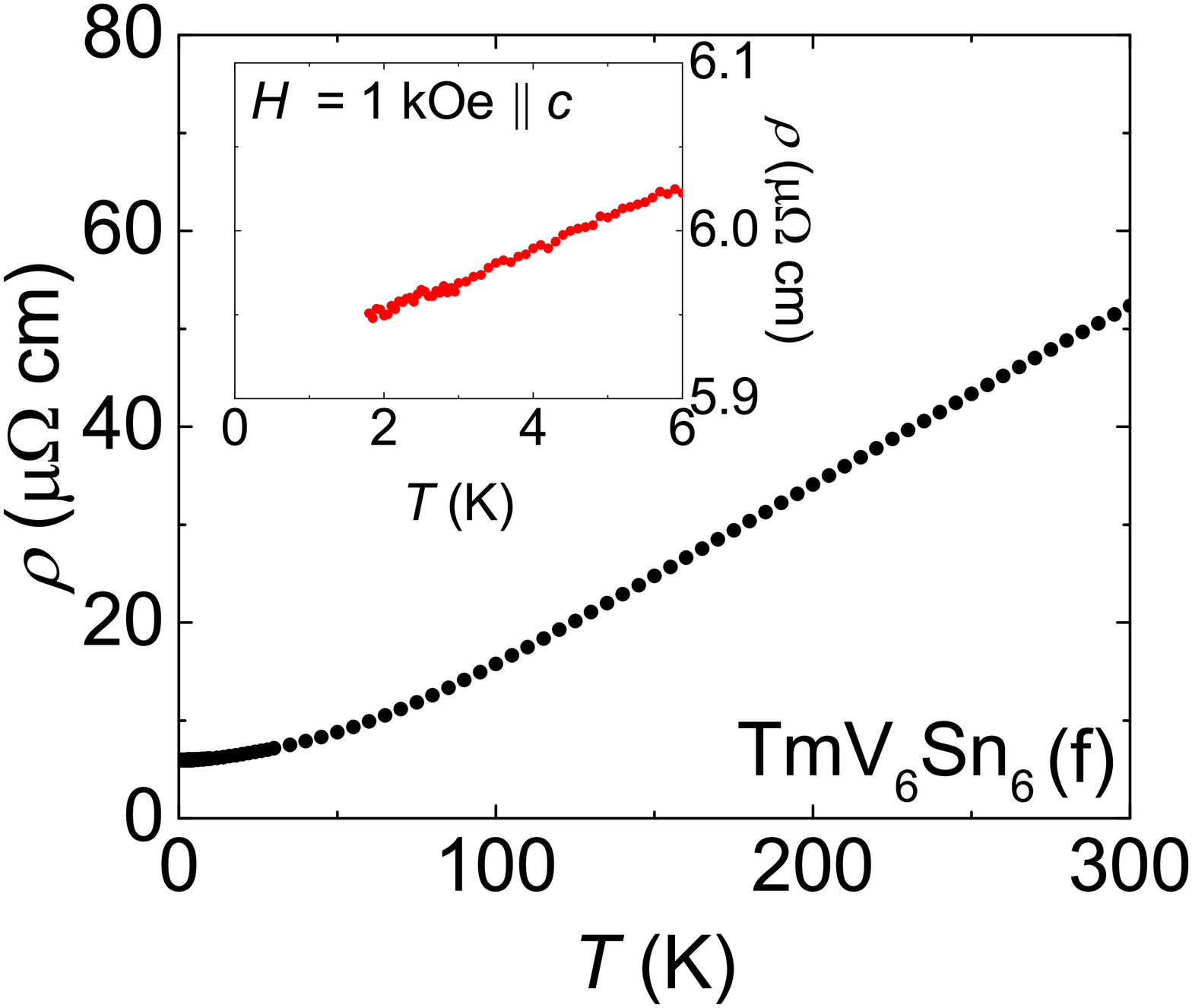}
\caption{Electrical resistivity, $\rho(T)$, curves of $R$V$_6$Sn$_6$ ($R$ = Gd - Tm). Insets show $\rho(T)$ at low temperatures, where vertical arrows indicate the magnetic ordering temperatures. For $R$ = Gd, Tb, and Tm, $\rho(T)$ at $H$ = 1 kOe is presented.}
\label{Fig6}
\end{figure}

Figures~\ref{Fig6} (a)$-$(f) present the $\rho(T)$ curves of $R$V$_6$Sn$_6$. Note that  due to the superconducting impurity phase $\rho(T)$ curves at $H$ = 1~kOe for $R$ = Gd, Tb, and Tm are plotted. At high temperatures, $\rho(T)$ follows typical metallic behavior with resistivity values ranging $10-100$~$\mu\Omega$~cm at 300~K. At low temperatures, $\rho(T)$ curves for $R$ = Gd - Ho show a sharp drop at $T_{N}$ due to the loss of spin disorder scattering, indicated by arrows in the insets of Fig.~\ref{Fig6} (a)$-$(d). $\rho(T)$ of GdV$_6$Sn$_6$ and YV$_6$Sn$_6$ is similar to that of previous reports. \cite{Ishikawa2021,Pokharel2021}

%Specific heat

\begin{figure}
\centering
\includegraphics[width=1\linewidth]{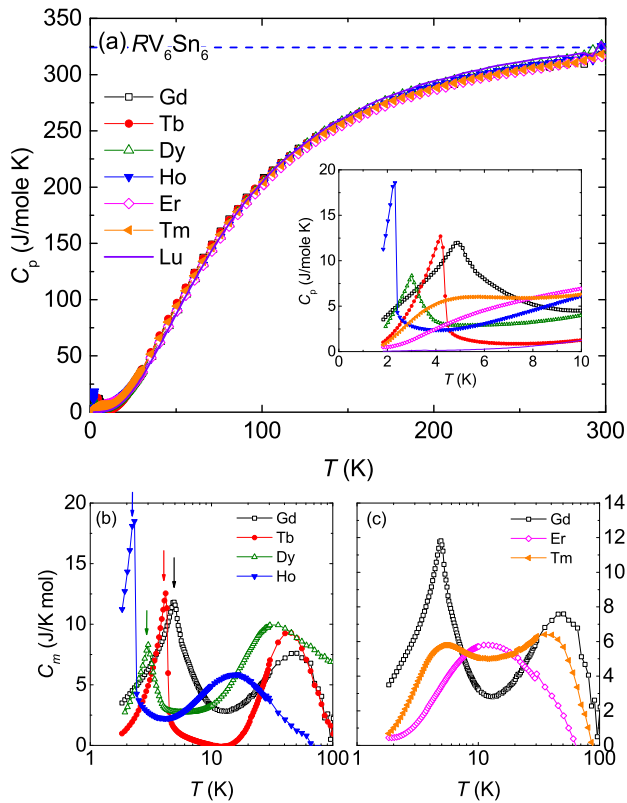}
\caption{(a) Specific heat $C_{p}$ curves of $R$V$_6$Sn$_6$ ($R$ = Gd - Tm and Lu). Inset shows $C_{p}$ below 10 K. (b) Magnetic part of the specific heat $C_{m}$ for $R$ = Gd - Ho. Vertical arrows indicate the magnetic ordering temperatures. (c) $C_{m}$ for $R$ = Gd, Er, and Tm.}
\label{Fig7}
\end{figure}

Figure~\ref{Fig7} (a) shows the $C_{p}(T)$ curves for $R$ = Gd - Tm and Lu. The specific heat of all $R$V$_6$Sn$_6$ compounds reaches the value close to the Dulong-Petit limit at 300~K. At low temperatures, specific heat measurements for $R$ = Gd - Ho show $\lambda$-like peaks (insets) as signatures of magnetic ordering, which are better seen in the magnetic part of specific heat ($C_{m}$). The magnetic ordering temperatures, determined from the peak positions, for $R$ = Gd, Tb, Dy, and Ho are $T_{N}$ = 4.9, 4.2, 3.0, and 2.4~K, respectively, which are consistent with magnetic susceptibility and resistivity measurements. For $R$ = Er and Tm, no peak in $C_{p}(T)$ is observed down to 1.8~K. The low-temperature $C_{m}$ curves for $R$ = Gd - Tm are obtained by subtracting the specific heat of LuV$_{6}$Sn$_{6}$ and plotted in Figs.~\ref{Fig7} (b) and (c). In addition to the sharp peaks at $T_{N}$, $C_{m}$ curves show broad maxima at higher temperatures. These maxima correspond to the Schottky contributions, as the $R^{3+}$ ions ($R$ = Tb - Tm) are influenced by the CEF. It has to be noted that $C_{m}$ of GdV$_{6}$Sn$_{6}$ shows an unusual Schottky-like anomaly (a broad maximum centered around 50~K), despite CEF splitting not being expected in Gd-based compounds. This anomaly may be due to the subtraction error. 

%4f moment, magnetic ordering, anisotropy

\begin{figure}
\centering
\includegraphics[width=1\linewidth]{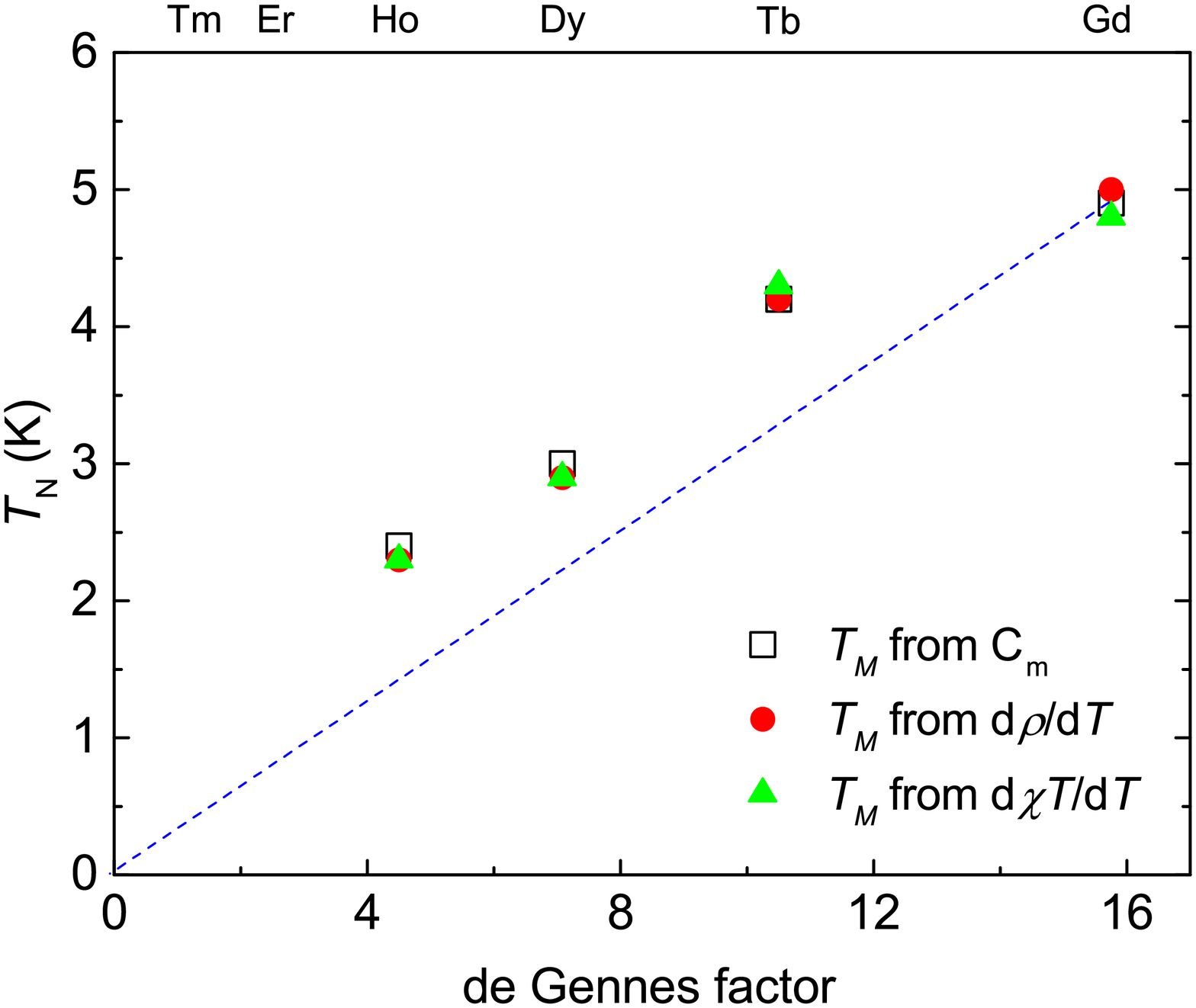}
\caption{Magnetic ordering temperature, $T_{N}$, as a function of de Gennes factor. The blue dashed line represents expected ordering temperatures for $R$ = Tb - Tm without CEF.}
\label{FIG8}
\end{figure}

The observed magnetic ordering temperatures for $R$V$_6$Sn$_6$ are plotted in Fig.~\ref{FIG8} as a function of de Gennes factor. Unlike $R$Fe$_6$Ge$_6$ \cite{Cadogan2001} and $R$Mn$_6$$X_6$ ($X$ = Ge and Sn) \cite{Venturini1991, Brabers1993} which have magnetic ordering temperatures above the room temperature, $R$V$_6$Sn$_6$ compounds indicate the magnetic ordering at much lower temperatures. When the transition metals possess no magnetic moments in this family of materials such as $R$Co$_6$X$_6$ ($X$ = Ge and Sn) \cite{Szytula2004} and $R$Cr$_6$X$_6$ \cite{Brabers1994}, a relatively low magnetic ordering temperature has been detected. Since the V ions possess no magnetic moments, the observed magnetic ordering is solely based on 4$f$ moments, and thus a very low magnetic ordering temperature is seen. 

When rare-earth ions are the only source of magnetism in an intermetallic compound, the magnetic ordering can be explained by indirect RKKY exchange interaction. In the molecular field approximation, the ordering temperature is proportional to de Gennes factor $(g_{J}-1)^{2}J(J+1)$ and can be defined as $T_N = 2/3\mathcal{J}(g_{J}-1)^2J(J+1)$, where $\mathcal{J}$ is the exchange parameter, $g$ is the Land\'e $g$ factor and $J$ is the total angular momentum quantum number of Hund's rule ground state of $R^{3+}$ ions. \cite{De1962,Noakes1982} For the compounds with heavy rare-earths, $T_N$ should decreases monotonically as $R$ traverses from Gd to Yb. The ordering temperatures in many rare-earth-based intermetallic compounds follow this scaling \cite{Morosan2005, Falkowski2007,Szytula2010,Marcinkova2015}. However, when there are strong CEF effects, a deviation from the linear de Gennes scaling has been observed for $R$ = Tb - Yb. \cite{Noakes1982, Van2007}. As shown in Fig.~\ref{FIG8}, the magnetic ordering temperature of $R$V$_6$Sn$_6$ indicates a slight deviation from the de Gennes scaling, expected to be due to the CEF effects. The observed anisotropic magnetic susceptibility and magnetization isotherm and broad maxima in specific heat clearly reflect the CEF effects on $R$V$_6$Sn$_6$ compounds. In addition, it has been shown a switch of easy magnetization direction from $c$-axis for $R$ = Tb, Dy, and Ho to $ab$-plane for $R$ = Er and Tm in many tetragonal and hexagonal rare-earth-based intermetallic compounds \cite{Noakes1982,Myers1999,Bud1999}. Typically, due to the change in sign of leading crystal field parameter, the switch of easy magnetization direction occurs between Ho and Er. Based on the point charge model, the leading crystal field parameter $B_{2}^{0}$ in hexagonal symmetry can be obtained by $B^{0}_{2}=\frac{10(\theta_{p}^{ab}-\theta_{p}^{c})}{3(2J-1)(2J+3)}$. \cite{Wang1971,Boutron1973} The estimated $B^{0}_{2}$ values from the magnetic susceptibility results are summarized in Table~\ref{tab:table1}. In $R$V$_6$Sn$_6$, the $B^{0}_{2}$ changes a sign between Ho and Er, suggesting that the detected magnetic anisotropy is mainly associated with CEF. When the strong CEF exchange interaction is present, the magnetic ordering temperature depends on $B^{0}_{2}$ and the large value of $B^{0}_{2}$ gives rise to enhance the magnetic ordering temperature, which breaks the simple de Gennes picture \cite{Noakes1982}. In $R$V$_6$Sn$_6$, although $T_{N}$ for Tb - Ho is enhanced, the maximum ordering temperature occurs in GdV$_6$Sn$_6$ and $T_{N}$ still follows the monotonic decrease from Gd to Ho. This implies that CEF interaction to enhance $T_{N}$ for $R$ = Tb - Ho is not large enough to exceed $T_{N}$ for $R$ = Gd \cite{Noakes1982,Zhou2009}. The $B^{0}_{2}$ value of TbV$_6$Sn$_6$ is greatest among $c$-axis ordering $R$V$_6$Sn$_6$ compounds, however $B^{0}_{2} \sim -1.4$~K is much smaller than that of other rare-earth-based compounds showing the highest ordering temperature for $R$ = Tb \cite{Noakes1982,Zhou2009}.

%CEF ground state & possible frustration

\begin{figure}
\centering
\includegraphics[width=1\linewidth]{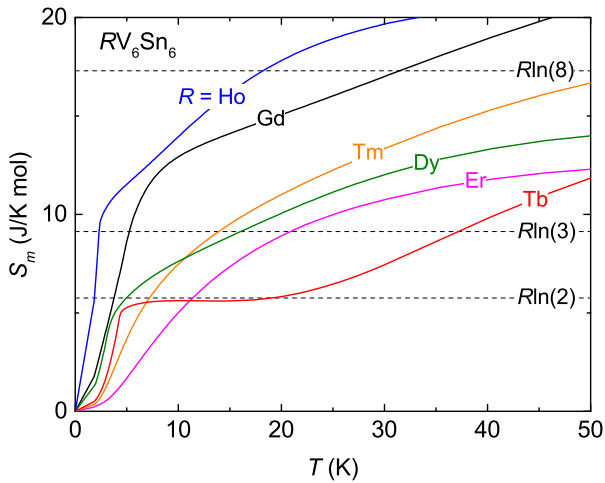}
\caption{Magnetic entropy, $S_{m}$, curves of $R$V$_6$Sn$_6$ ($R$ = Gd - Tm).}
\label{Fig9}
\end{figure}

The magnetic entropy, $S_{m}$, curves of $R$V$_6$Sn$_6$ are shown Fig.~\ref{Fig9}, where  $S_{m}$ is estimated by integrating $C_{m}/T$. Below 1.8~K, the missing entropy is estimated by assuming $C = 0$ at $T$ = 0~K. $S_{m}$ provides insight into the underlying CEF ground state of magnetic $R$V$_6$Sn$_6$. At $T_{N}$, $S_{m}$ of TbV$_{6}$Sn$_{6}$ reaches a value close to $R\ln(2)$, suggesting a possible pseudo-doublet magnetic ground state (non-Kramers doublet, a singlet ground state with a singlet excited state). For DyV$_{6}$Sn$_{6}$, the value of $S_{m}$ at $T_{N}$ suggests a Kramers doublet ground state. For HoV$_{6}$Sn$_{6}$, $S_{m}$ at $T_{N}$ reaches a value slightly above $R\ln(3)$, suggesting a possible triplet or pseudo-triplet ground state (a singlet ground state with either a doublet excited state or two more singlet excited states, all with close energies). Note that the ground state entropy must be properly confirmed by specific heat measurements below 1.8~K. Interestingly, only $\sim$50~\% of the $R\ln(8)$ magnetic entropy for GdV$_{6}$Sn$_{6}$ is recovered at $T_{N}$. In general, Gd$^{3+}$ ions in intermetallic compounds are in an $L = 0$ ($S$-state), no CEF effect is expected. Therefore, the full $R\ln(8)$ magnetic entropy is expected to be recovered at the ordering temperature. However, for GdV$_{6}$Sn$_{6}$ the full $R\ln(8)$ entropy is recovered at $\sim$30~K which is significantly higher than $T_{N}$ = 4.9~K. Obviously the entropy is overestimated at high temperatures, where the entropy continues to increase beyond $R\ln(8)$ above 30~K, suggesting a subtraction error in the estimate of $C_{m}$. Since the obtained effective moment and saturation moment for GdV$_6$Sn$_6$ are close to the theoretical Gd$^{3+}$ ion values, $C_{m}$ is purely based on Gd magnetic contributions. When it is assumed that GdV$_6$Sn$_6$ has no additional magnetic transitions below 1.8~K, the reduced entropy at $T_{N}$ cannot be simply related to the subtraction error. The reduced entropy may raise the possibility of a mixed magnetic structure, where both magnetic and non-magnetic sites coexist on an equivalent crystallographic site. This kind of magnetic structure has been evident in geometrically frustrated metallic systems \cite{Nakamura1999, Mentink1994, Stockert2020}. It has also been shown theoretically and experimentally that the RKKY interaction in two dimensional hexagonal lattice gives rise to complex spin structures and potentially hosts a skyrmion phase in conjunction with geometrical frustration \cite{Kurumaji2019,Zhang2020,Tokura2020}. A recent study on GdV$_6$Sn$_6$ has suggested that Gd ions form noncollinear spin structure below the magnetic ordering temperature \cite{Pokharel2021}.

Peculiar electronic states have been observed in Fe$_3$Sn$_2$ \cite{Fenner2009, Kida2011, Li2019}, FeSn \cite{Kang2020}, and Co$_3$Sn$_2$S$_2$ \cite{Wang2018}, where the moment-bearing transition metals form a Kagome lattice. Recently $A$V$_3$Sb$_5$ ($A$ = K, Rb Cs) systems have shown charge density wave (CDW), superconductivity, and anomalous Hall effect \cite{Ortiz2020,Ortiz2021,Yin2021} all of which have opened up a new direction for exploring unconventional electronic properties associated with the non-magnetic Kagome layer. The V ions in $R$V$_6$Sn$_6$ possesses no magnetic moment and form the Kagome lattice. In particular, the Kagome layer in these compounds contain no other atoms in the quasi two dimensional Kagome network. Since $R$V$_6$Sn$_6$ ($R$ = Gd, Y, and Sc) compounds showed non-linear Hall resistivity, topologically nontrivial band structure, and CDW \cite{Ishikawa2021,Pokharel2021,Arachchige2022}, it is of great interest to check other rare-earth-based compounds ($R$ = Tb - Tm and Lu) whether the unconventional electronic states exist in the whole family of materials.

\section{Summary}

Single crystals of $R$V$_6$Sn$_6$ ($R$ = Y, Gd - Tm, Lu) are grown by Sn-flux and their physical properties are investigated by magnetization, specific heat, and resistivity measurements. Powder X-ray diffraction patterns reveal that these compounds crystallize into the HfFe$_6$Ge$_6$-type structure, where V ions form a well isolated Kagome layer and rare-earth ions occupy on a well separated triangular lattice. Due to the CEF effects on rare-earth ions the magnetization as a function of temperature and magnetic field shows a large magnetic anisotropy, where the easy magnetization direction is $c$-axis for $R$ = Tb - Ho and $ab$-plane for $R$ = Er and Tm. At low temperatures, the antiferromagnetic ordering is observed below 5~K for $R$ = Gd - Ho and no magnetic ordering is observed down to 1.8~K for $R$ = Er and Tm. Since V ions possesses no magnetic moments, the magnetic properties of intermetallic $R$V$_6$Sn$_6$ compounds are solely governed by localized 4$f$ electron moments.

\section{Acknowledgments}
This work was supported by the Canada Research Chairs, Natural Sciences and Engineering Research Council of Canada, and Canada Foundation for Innovation program.

%\newpage{\pagestyle{empty}\cleardoublepage}

\end{document}